\documentclass[12pt,preprint,graphicx]{aastex}

\slugcomment{}

\lefthead{Hachisuka et al.}
\righthead{The Outer Spiral Arm}

\def\uas {$\mu$as}
\def\kms {km~s$^{-1}$}
\def\Vlsr {$V_{\rm {LSR}}$}
\def\masy {mas~yr$^{-1}$}

\def\Us   {$U_s$}

\def\Ws   {$W_s$}

\def\etal {et al.~}

\def\HII  {H~{\footnotesize II}}

\def\Vsbar {\ifmmode {\overline{V_s}}\else {$\overline{V_s}$}\fi}
\def\Usbar {\ifmmode {\overline{U_s}}\else {$\overline{U_s}$}\fi}
\def\Wsbar {\ifmmode {\overline{W_s}}\else {$\overline{W_s}$}\fi}

\def\d    {\ifmmode {{\rlap{.}}^\circ}\else {${\rlap{.}}^\circ$}\fi}
\def\s    {\ifmmode {{\rlap{.}}^s}\else {${\rlap{.}}^s$}\fi}
\def\as   {\ifmmode {{\rlap{.}}^{''}}\else {${\rlap{.}}^{''}$}\fi}

\newbox\grsign \setbox\grsign=\hbox{$>$} \newdimen\grdimen \grdimen=\ht\grsign
\newbox\laxbox \newbox\gaxbox
\setbox\gaxbox=\hbox{\raise.5ex\hbox{$>$}\llap
     {\lower.5ex\hbox{$\sim$}}}\ht1=\grdimen\dp1=0pt
\setbox\laxbox=\hbox{\raise.5ex\hbox{$<$}\llap
     {\lower.5ex\hbox{$\sim$}}}\ht2=\grdimen\dp2=0pt
\def\gax{\mathrel{\copy\gaxbox}}

\begin{document}

\title{Parallaxes of Star Forming Regions in the Outer Spiral Arm of the Milky Way}
   \author{K. Hachisuka\altaffilmark{1,2,3},
           Y. K. Choi\altaffilmark{4,6},
	   M. J. Reid\altaffilmark{5},
           A. Brunthaler\altaffilmark{6},
           K. M. Menten\altaffilmark{6},
	   A. Sanna\altaffilmark{6}, \\
	   T. M. Dame\altaffilmark{5}
}

\altaffiltext{1}{Research Institute for Time Studies, Yamaguchi University, 
                      Yoshida 1677-1, Yamaguchi, Yamaguchi, 753-8511, Japan}
\altaffiltext{2}{Shanghai Astronomical Observatory, Chinese Academy of 
                 Science, 80 Nandan Road, Shanghai, 200030, China}
\altaffiltext{3}{Center for Astronomy, Ibaraki University, 2-1-1 Bunkyo, Mito, Ibaraki 210-8512, Japan}
\altaffiltext{4}{Korea Astronomy and Space Institute, Daedeokdae-ro 776, 
                 Yuseong-gu, Daejeon, 305-348, Korea}
\altaffiltext{5}{Harvard-Smithsonian Center for Astrophysics,
              60 Garden Street, Cambridge, MA 02138, USA}
\altaffiltext{6}{Max-Planck-Institut f\"ur Radioastronomie,
                  Auf dem H\"ugel 69, D-53121, Bonn, Germany}

\begin{abstract}
We report parallaxes and proper motions of three water maser sources in
high-mass star-forming regions in the Outer Spiral Arm of the Milky Way.
The observations were conducted with the Very Long Baseline Array as  
part of Bar and Spiral Structure Legacy Survey and double the number of 
such measurements in the literature.  
The Outer Arm has a pitch angle of $14\d9 \pm 2\d7$ and a Galactocentric 
distance of $14.1 \pm 0.6$ kpc toward the Galactic anticenter.  The average 
motion of these sources toward the Galactic center is $10.7 \pm 2.1$ 
\kms\ and we see no sign of a significant fall in the rotation curve out to 15 kpc 
from the Galactic center.  The three-dimensional locations of these star-forming 
regions are consistent with a Galactic warp of several hundred parsecs from the plane.
\end{abstract}


\section{Introduction}

Looking outward from the Sun in the direction of the Galactic anticenter, one's
sight line passes through the Local, Perseus and Outer spiral arms of the Milky Way
(e.g. Xu et al. 2013; Choi et al. 2014; Dame et al. 2001; Vall{\'e}e 2005;  
Churchwell et al. 2009; Reid et al. 2014).The Outer Arm may originate near the 
Galactic bar (McClure-Griffiths et al. 2004; Nakanishi \& Sofue 2006;  Levine et al. 
2006; Dame \& Thaddeus 2011) and wind its way outward through more 
than 360 deg of Galactocentric azimuth, until reaching and then passing the 
anticenter direction.  While, for example, there is prolific star formation in the 
Perseus arm, there is relatively little activity in the Outer Arm in the second and 
third quadrants.  This makes it challenging to accurately trace its structure and 
determine kinematic properties.  

Trigonometric parallax distances, $D_\pi$, have been measured for only three 22 
GHz water maser sources in the Outer Arm: $D_\pi=9.25\pm0.43$ kpc for G075.30+01.32 
(Sanna et al. 2012), $D_\pi=5.99\pm0.22$ kpc for WB 89--437 (G135.27+02.79) 
(Hachisuka et al. 2009), and $D_\pi=5.28\pm0.23$ kpc (Honma et al. 2007) or 
$D_\pi=4.05\pm0.65$ kpc (Asaki et al. 2014) for S 269 (G196.45--01.67).  
While clearly more astrometric data are desirable in order to better understand 
the properties of the Outer Arm, few maser sources have been discovered, 
owing to the low star formation rate and general weakness of these masers 
(e.g. Wouterloot et al. 1993; Szymczak \& Kus 2000). 

Astrometry with Very Long Baseline Interferometry (VLBI) can yield trigonometric 
parallaxes with accuracies better than $\pm10$ \uas\ and is an excellent tool 
to study Galactic structure and dynamics (e.g. Reid et al. 2009b; Honma et al 2012; 
Reid et al. 2014).  In this paper, we report parallaxes and proper motions of three
water masers associated with high-mass star-forming regions (HMSFRs):
G$097.53+03.18$, G$168.06+00.82$ and G$182.67-03.26$.  These measurements
are part of the Bar and Spiral Structure Legacy (BeSSeL) Survey and
double the number of sources with accurate parallaxes in the Outer Arm.

\section{Observations}

We used the National Radio Astronomy Observatory's (NRAO)
\footnote{The National Radio Astronomy Observatory is a facility of the National Science 
Foundation operated under cooperative agreement by Associated Universities, Inc.} 
Very Long Baseline Array (VLBA) to observe 22 GHz water maser sources under programs
BR145G, H, and V.   We observed with four adjacent 8 MHz bands in right and left
circular polarization, with the second band centered on the water maser.  
The continuum and line data were correlated with the VLBA DiFX \footnote{DiFX, a software 
Correlator for VLBI, is developed as part of the Australian Major National Research 
Facilities Programme by the Swinburne University of Technology and operated under license.},
producing 16 and 256 spectral channels per band, respectively.  
Additionally, we placed four ``geodetic blocks'' throughout each seven hour track, 
as described in Reid \etal (2009a), in order to measure and remove residual tropospheric 
delays relative to the correlator model.
The parallax observations involved rapid switching between compact extragalactic 
continuum sources and water maser sources.  For each maser source, we
observed at six epochs spanning one year (see Tables~\ref{table:1} and 
\ref{table:1-2} for details), with a sequence designed to yield 
good parallax results for a maser spot that might last only seven months.
Data reduction was performed using the NRAO Astronomical Image 
Processing System, as described in Reid et al. (2009a). 

The water masers were detected at all epochs, and generally we chose a 
strong, compact maser spot to use as the interferometer phase reference. 
The second continuum source associated with G$182.68-03.26$,
while detected at the first epoch, was not useful for parallax results owing to
its large angular separation ($3^\circ$) from the maser.

\begin{table*}[h]
\caption{Source Information.\label{table:1}}
\begin{center}
\begin{tabular}{rcllcccc}
\hline \hline
\multicolumn{1}{c}{Source} & Type & R.A. (J2000) & Decl. (J2000) & $\theta_{sep}$ & P.A. &Brightness & 
N \\
& & (h \ \ m \ \ s)  & ($^\circ$ \ \  ' \ \  ") & ($^\circ$)     & ($^\circ$) & (mJy beam$\mbox{}^{-1}$) & \\
\hline
G097.53+03.18  &M& 21:32:12.4343  & +55:53:49.689  & --  & --  & -- & 11 \\
J2127+5528   &C& 21:27:32.27527 & +55:28:33.9771 & 0.78 & -122 & 11 \\
J2139+5540   &C& 21:39:32.61754 & +55:40:31.7711 & 1.05 & \ 101& 16 \\
J2123+5500   &C& 21:23:05.31350 & +55:00:27.3250 & 1.57 & -124 & 94 \\
J2117+5431   &C& 21:17:56.48440 & +54:31:32.5030 & 2.45 & -123 & 92 \\
\hline
G168.06+00.82&M& 05:17:13.7436  & +39:22:19.915  & --  & --  & -- & 4 \\
J0523+3926   &C& 05:23:51.2364  & +39:26:57.736  & 1.28 & \ 86 & 17 \\
J0509+3951   &C& 05:09:48.8173  & +39:51:54.618  & 1.51 & -70 & 69  \\
J0512+4041   &C& 05:12:52.5428  & +40:41:43.620  & 1.56 & -32 & 224 \\
\hline
G182.68--03.26&M& 05:39:28.4248  & +24:56:31.946  & --  & --  & -- & 4 \\
J0540+2507   &C& 05:40:14.3428  & +25:07:55.349  & 0.26 & \ 42 & 12 \\ 
J0550+2326   &C& 05:50:47.3909  & +23:26:48.177  & 2.98 & 120 & 33 \\
\hline
\end{tabular} \\
\end{center}
{\footnotesize {\bf Notes.} Type: M for maser source; C for continuum source;
$\theta_{sep}$ and P.A. are angular separation and position angle 
east of north from each maser source.  N is number of maser spots for  
astrometric fitting. }
\end{table*}

\begin{table*}[h]
\caption{Observation Epochs\label{table:1-2}}
\begin{center}
\scriptsize
\begin{tabular}{lcccccc}
\hline \hline
Source & \multicolumn{6}{c}{Epoch} \\
 & 1st & 2nd & 3rd & 4th & 5th & 6th \\ 
\hline
G097.53$+$03.18 & 2010/Dec/10&2011/Feb/21&2011/Apr/30&2011/May/29&2011/Jul/17&2011/Nov/17 \\
G168.06$+$00.82 & 2010/Apr/16&2010/Jun/17&2010/Aug/15&2010/Sep/24&2010/Nov/15&2011/Mar/7 \\
G182.68$-$03.26 & 2010/Apr/23&2010/Jun/20&2010/Aug/23&2010/Sep/25&2010/Nov/16&2011/Mar/21 \\
\hline
\end{tabular} \\
\end{center}
\end{table*}

\section{Astrometric Results}

\begin{table*}[h]
\begin{center}
\caption{Parallax and Proper Motions}
\begin{tabular}{lccccc}
\hline \hline
Name & $\pi$ & $\mu_{\mbox{x}}$  & $\mu_{\mbox{y}}$  &\Vlsr     \\
     & (mas) & (\uas)  & (\uas) & (\kms) \\             
\hline
G097.53+03.18 & 0.133$\pm$ 0.017 & --2.94$\pm$ 0.06 &  --2.48$\pm$ 0.14 & --75$\pm$  3 \\
G168.06+00.82 & 0.201$\pm$ 0.024 &   0.58$\pm$ 0.31 &  --0.78$\pm$ 0.31 & --28$\pm$  5 \\
G182.67-03.26 & 0.157$\pm$ 0.042 &   0.44$\pm$ 0.34 &  --0.45$\pm$ 0.35 &  
--8 $\pm$10 \\
\hline
\end{tabular}
\end{center}
{\footnotesize
{\bf Notes.} Columns 1 through 5 give the source name, annual parallax ($\pi$), 
absolute proper motion in the eastward  ($\mu_{\mbox{x}}$) and northward 
($\mu_{\mbox{y}}$) directions, and the mean Local Standard of Rest velocity (\Vlsr).
The uncertainty assigned to \Vlsr\ is the difference between mean velocity of masers and 
of thermal CO emission from the parent molecular cloud, except for G182.67-03.26 
where there is no information for thermal molecular lines and we adopt an
uncertainty of 10 \kms\ (Wouterloot et al. 1995).  Other characteristics of these 
sources can be found Section 3.1, 3.2 and 3.3.}
  \label{table:2}
\end{table*}

Source positions were estimated by fitting an elliptical Gaussian brightness
distribution to the interferometric images.  These fits produced formal
position uncertainties, which do not include systematic sources of error,
typically dominated by uncompensated atmospheric delays.  Therefore, during the 
fitting process, we included``error floors'' to account for these systematic 
uncertainties. The error floors were added in quadrature to the formal position
uncertainties and adjusted until we achieved $\chi_\nu^2$ per degree of 
freedom values of near unity in each coordinate.  The parallax and absolute proper 
motion results are shown in Table~\ref{table:2}.  When we report an annual 
parallax by combining two or more maser spots, we multiplied the formal fitting error 
by $\sqrt{N}$, where $N$ is the number of maser spots, 
conservatively allowing for the possibility of 100\% correlated position uncertainties 
among the spots.  Throughout this paper we adopt customary
definitions for a maser spot (the emission in one spectral channel)
and maser feature (typically several channels that comprise a Gaussian line shape).
On the one hand, only the most compact maser spots, detected at all epochs,
were used to fit the parallax curve. On the other hand, in order to estimate
the Galactic proper motion of the star-forming region, we took into account
the internal motions of as many maser spots as possible. Their average motion
has been taken as representative of the motion of the central star and was
used to correct the proper motion value inferred with the parallax fitting (Table 3).

Water masers are typically associated with outflowing motions from young stellar 
objects tracing velocities on the order of tens of \kms.  
For the line-of-sight velocity component, we rely on emission from thermal
lines of CO and other molecules, which sample larger portions of the
molecular cloud, to provide a more robust estimate of the average motion of 
the central star.

\subsection {G097.53+03.18}

This region, also known as S~128, consists of a diffuse and compact \HII\ region.  
The 22 GHz water masers are located near the northern compact \HII\ region 
(S~128N; Haschick \& Ho ).  There are also two water maser spots separated by 
about 9$\farcs$ from S~128N,  (Haschick \& Ho 1985), but these were not imaged owing 
to fringe-rate smearing. The LSR velocity for CO emission is $-74$ \kms\ 
(Haschick \& Ho 1985) and for CS emissionis $-70$ \kms\ (Plume et al. 1997).  
The water masers have LSR velocitiesfrom $-69$ to $-83$ \kms, with the peak 
brightness at $-77$ \kms. 

\begin{figure}[ht]
\resizebox{\hsize}{!}{\includegraphics[angle=0]{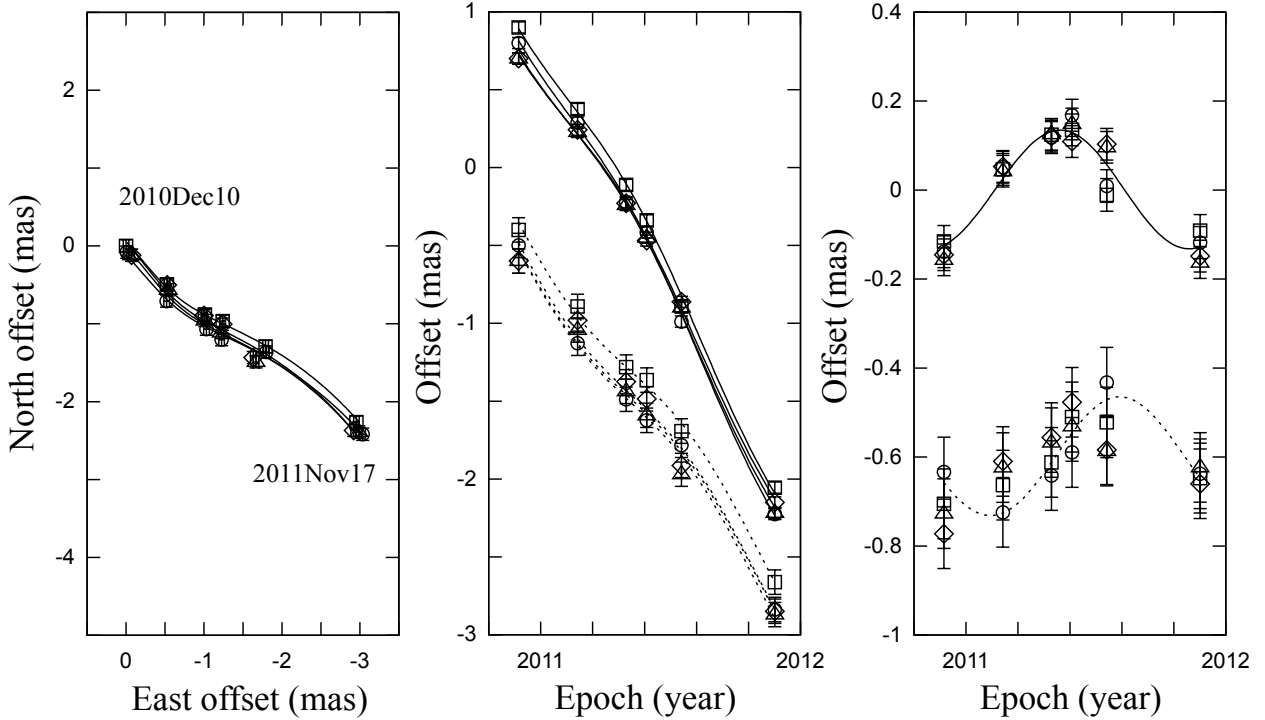}}
\caption{
{\footnotesize 
Parallax and proper motion data for the maser spot at \Vlsr$=-77.0$ \kms\   
toward G097.53+03.18 relative to the background sources J2127+5528 (squares), 
J2139+5540 (circles), J2123+5500 (triangles) and J2117+5431 (diamonds). 
{\it Left panel:} position offsets on the sky.
{\it Center panel:} offsets of maser spots eastward ({\it solid line}) and northward 
({\it dashed line}) vs. time. 
{\it Right panel:} same as the center panel, but with the proper motion removed. 
}
}
\label{astrometry-g097}
\end{figure}

\begin{figure}[ht]
\resizebox{\hsize}{!}{\includegraphics[angle=0]{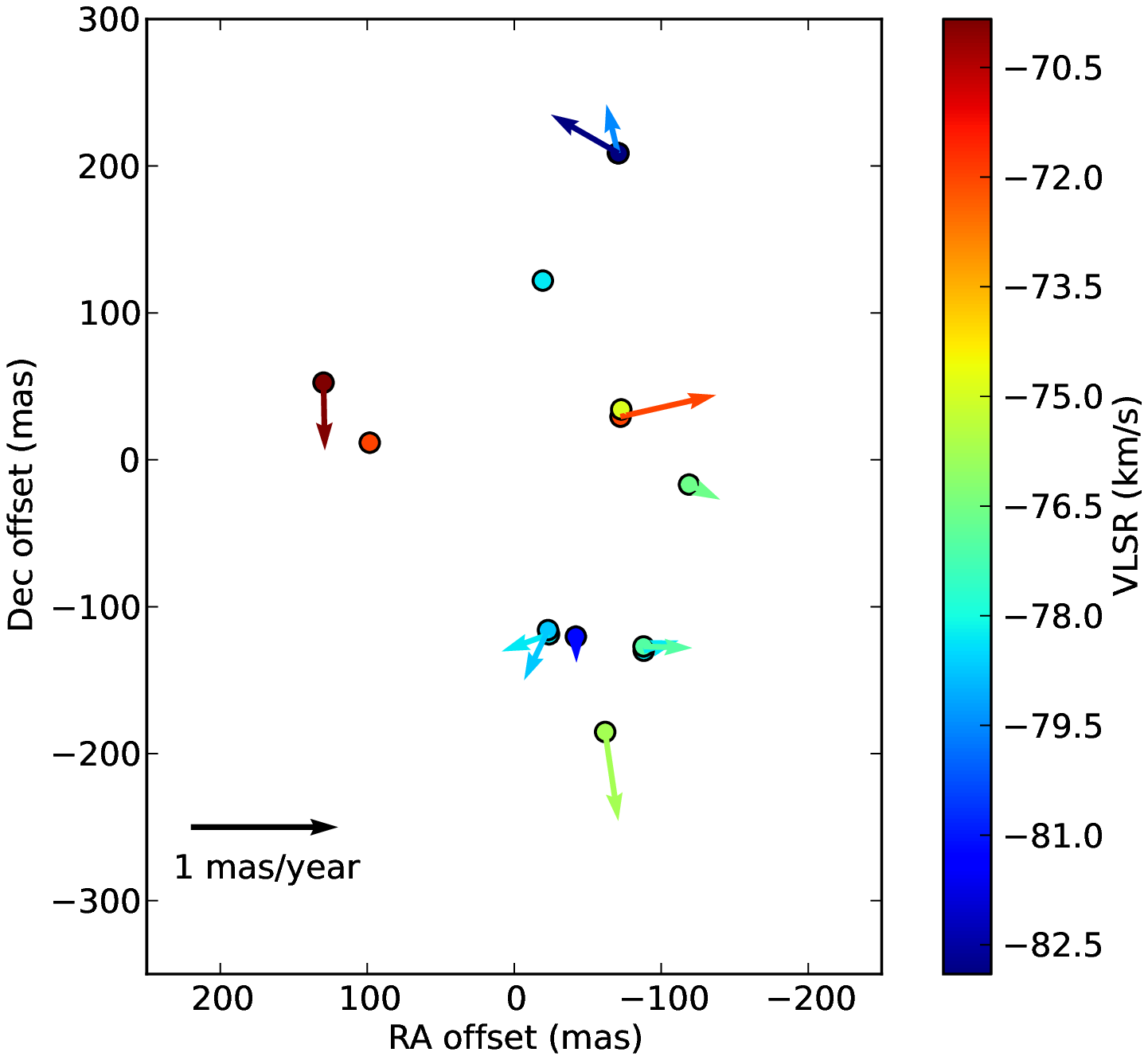}}
\caption{
{\footnotesize 
Proper motions of water masers toward G097.53+03.18.  
Vectors indicate motions about the estimated center of expansion
near (0,0) mas.  Spots without vectors indicate that the maser was detected 
only at one or two epochs.
}
}
\label{internal_g097}
\end{figure}

We choose 11 maser spots from seven different maser features for the parallax and proper 
motion fitting.  The results for the maser spots and background continuum sources are 
listed in Tables~\ref{table:g097a}.
The parallax and proper motion data for the maser spot at \Vlsr$=-77.0$ \kms\ 
are shown in Figure~\ref{astrometry-g097}.
The distribution of maser features is complex we fitted a uniformly 
expanding source model (Sato et al. 2010), giving the results shown in 
Figure~\ref{internal_g097}.  On average, maser features are expanding slowly 
($2.2\pm5.3$ \kms) and the average proper motion of the central exciting star(s) 
is $0.33\pm0.15$ \masy\ eastward and $0.01\pm0.15$ \masy\ northward.

\subsection {G168.06+00.82}

This source is also known as Mol~8 (Brand et al. 2001).   
The 22 GHz water masers are located toward the peak emission from 
thermal molecular lines (Brand et al. 2001) and millimeter-wavelength dust continuum 
(Molinari et al. 2000). The LSR velocity peaks for HCO$^{+}$, $\mbox{}^{13}$CO and 
CS are all near $-25$ \kms.  The radial velocity range of the water masers is
small, from $-25$ to $-31$ \kms, reasonably close to the thermal line peaks.

\begin{figure}[ht]
\resizebox{\hsize}{!}{\includegraphics[angle=0]{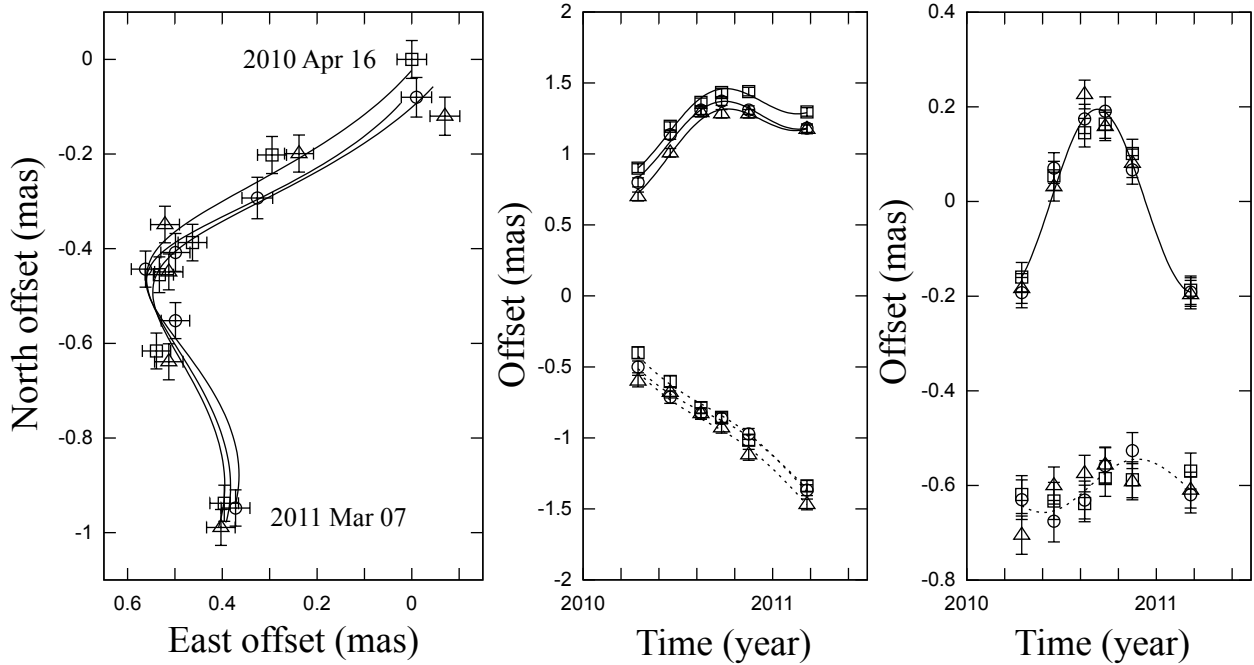}}
\caption{
{\footnotesize
Annual parallax and proper motion of the maser spot at \Vlsr=$-28.74$ \kms\ 
toward G168.06+00.82 relative to background source, J0512+4041 (squares), 
 J0523+3926 (circles) and J0509+3951 (triangles). 
See Figure~\ref{astrometry-g097} caption for details. 
}
}
\label{astrometry-g168}
\end{figure}

Several maser features were detected at all epochs and we choose four maser spots 
from one feature and one maser spot from other feature for astrometric 
fitting (see Table~\ref{table:g168}). Other maser spots in other features were not used 
for the astrometic fitting because they displayed complex, time varying blended structures. 
The combined annual parallax and average absolute proper motion are shown in Table~\ref{table:2}. 
The parallax and proper motion fit for the maser spot at an LSR velocity of 
$-28.74$ \kms\ is shown in Figure~\ref{astrometry-g168}.

Honma et al. (2011) measured the parallax and proper motion of the
water masers in this source (also known as IRAS~$05137+3919$).  Their
parallax is based on two maser features (one feature detected in five
spectral channels), neither detected over a time span longer than 
$\approx0.6$ yr.  Honma et al. favor a weighted parallax of $0.086\pm0.027$ mas.
If one adopts their solution with equal weights for the two features 
and allows for the parallax estimates of these features being correlated
(as expected for atmospheric mis-modeling being the dominant source of
systematic uncertainty), then their parallax result would be
$0.103\pm0.030$ mas.   There is some tension with our result of
$0.201\pm0.024$ mas, as the difference between the two measurements
would be $0.098\pm0.039$ mas.

\begin{figure}[ht]
\resizebox{\hsize}{!}{\includegraphics[angle=0]{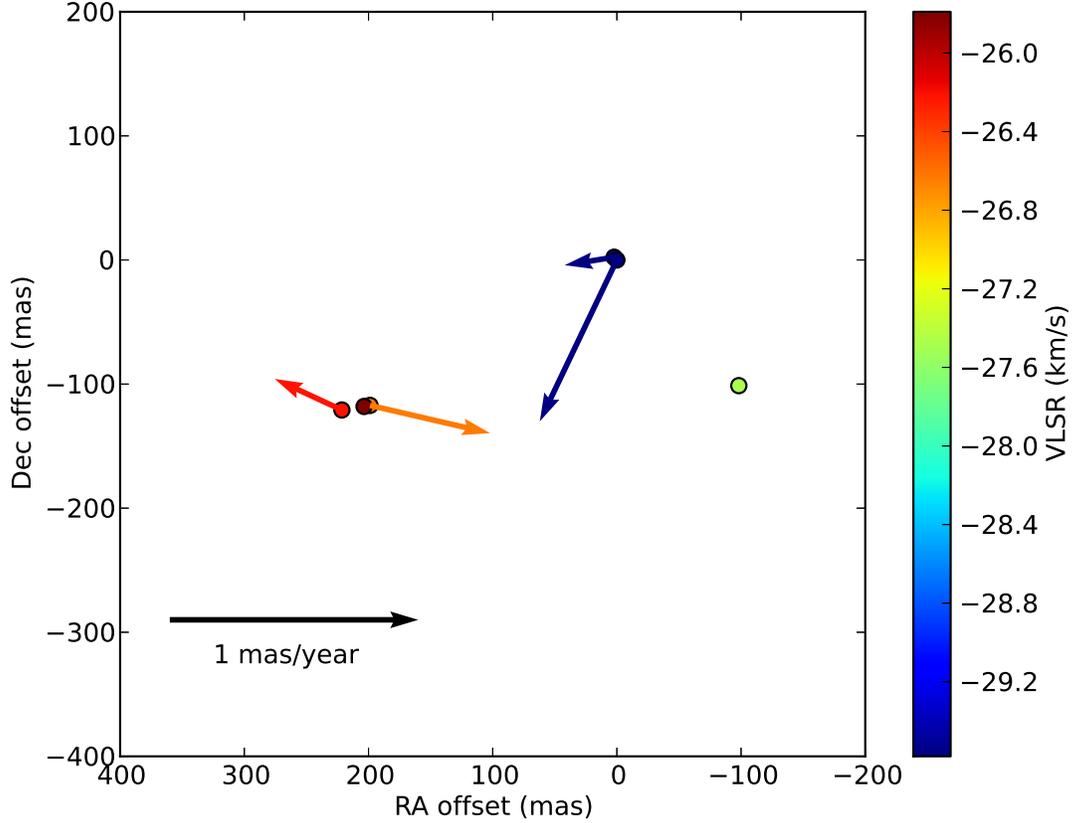}}
\caption{
{\footnotesize 
Proper motions of water masers toward G168.06+00.82.  
Vectors indicate motions about the estimated center of expansion
near (0,0) mas.  Spots without vectors indicate that the maser was detected only 
one or two epochs.
}
}
\label{internal_g168}
\end{figure}

Even though this source has only a small number of maser features, we did fit an 
expanding flow in order to estimate the central star's motion.  However, we 
conservatively assign an large uncertainty of $\pm7$ \kms\ ( $\sim$0.3 \masy\ at 5.0 kpc)
for the components of proper motion of the central star. 
The internal motions of maser features with the central star's estimated motion removed  
are shown in Figure~\ref{internal_g168}. 

\subsection {G182.67--03.26}

\begin{figure}[ht]
\resizebox{\hsize}{!}{\includegraphics[angle=0]{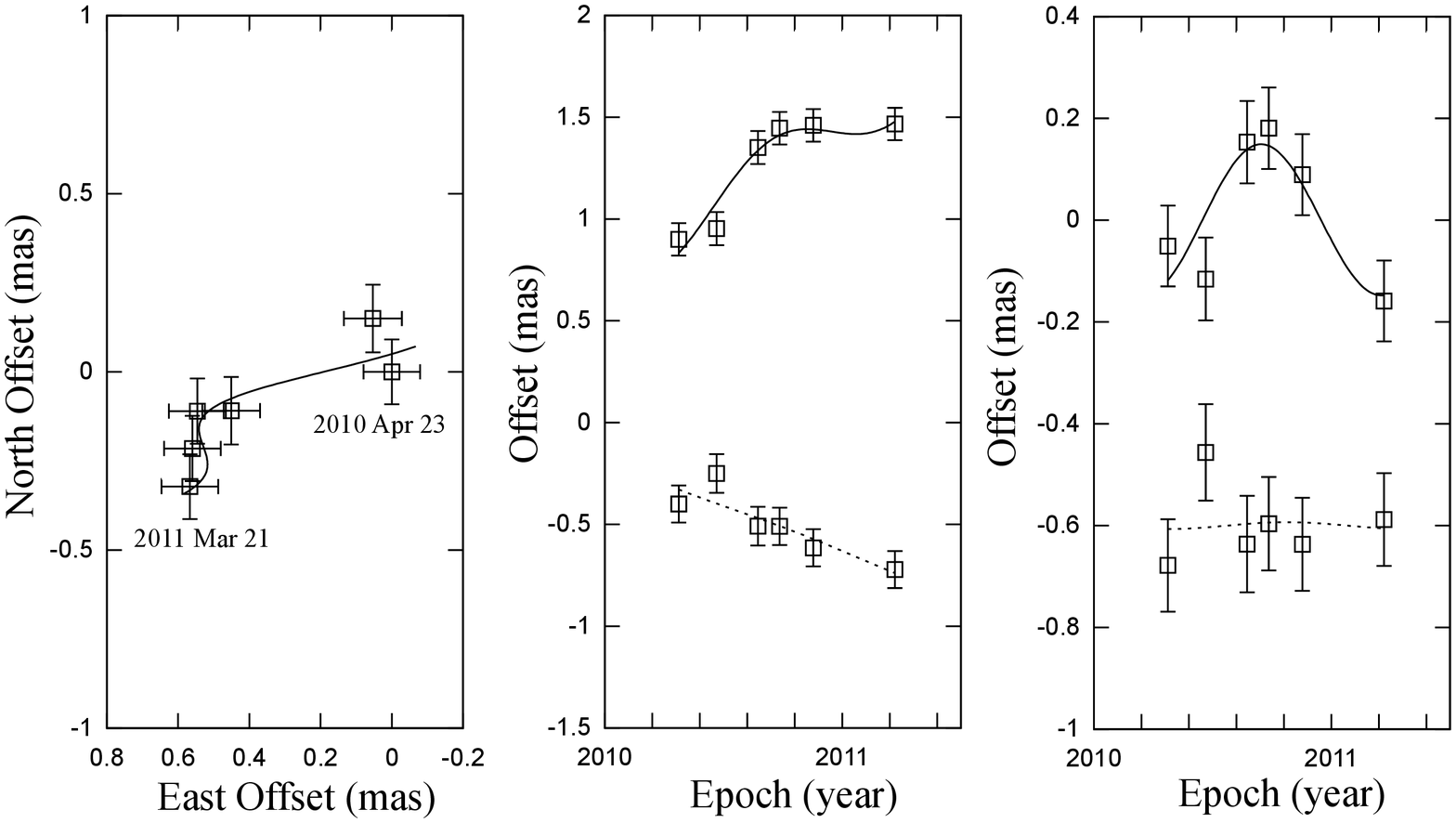}}
\caption{
{\footnotesize
Annual parallax and proper motion of the maser spot at \Vlsr=$-7.89$ \kms\ toward  
G182.67--03.26 relative to background source J0540+2507. 
See Figure~\ref{astrometry-g097} for details.
}
}
\label{astrometry-g182}
\end{figure}

We are unaware of measurements of thermal emission from molecular lines in 
the literature for this source.   The radial velocity range of the 22 GHz water masers 
is very small, $-5$ to $-9$ \kms.  Four maser spots at three different positions 
were detected at all epochs spanning one year, and we used these spots for the 
astrometric fitting (see Table~\ref{table:g182}).
The parallax and proper motion of the maser spot at \Vlsr$=-5.79$ \kms\ is 
shown in Figure~\ref{astrometry-g182}. The combined parallax for 
maser spots is $0.157\pm0.042$ mas, corresponding to a distance of 6.4 kpc.

Estimating the absolute motion of the central exciting star is difficult in this source,
since only three maser features were detected. 
The mean proper motion for the three maser features that could be traced for two or more
epochs was $0.35\pm0.23$ \masy\ eastward and $-0.14\pm0.47$ \masy\ northward, 
or $11\pm7$ \kms\ and $4\pm14$ \kms\ at the distance of 6.4 kpc.   
Owing to the minimal number of maser spots with measured motions,
we conservatively add ( 10 \kms )/( 6.7 kpc ) =0.33 \masy\ in quadrature with the
measurement uncertainty when transferring the maser motion to that of the 
central star.

\section{Discussion}

There are now VLBI astrometric observations toward six 22 GHz water maser 
sources in the outer spiral arm of the Milky Way (see Table \ref{table:4-1}).
The locations of these HMSFRs in the Galaxy and their 
peculiar motions relative to a Galactic rotation model are given in
Table~\ref{table:4-1}.   The peculiar motions assumed a flat rotation curve 
with $\Theta = 239 \pm 7$ \kms\ and the distance to the Galactic 
center of $R_{0} = 8.3 \pm 0.2$ kpc (Brunthaler et al. 2011), and 
the Solar motion of $(U_{s},V_{s},W_{s})$ = (11.10, 12.24, 7.25) \kms 
(Sch{\"o}nrich et al. 2010). 

\begin{table*}[h]
\caption{Outer Arm Sources.\label{table:4-1}}
\scriptsize
\begin{center}
\begin{tabular}{llllcccccc}
\hline \hline
Name & IRAS  & WB89 & Other & R$_{\mbox{GC}}$  & z &$U_{s}$ & $V_{s}$ & $W_{s}$ & Ref.     \\
     &       &      &Name   & (kpc) & (kpc) &(km s$^{-1}$) & (km s$^{-1}$) & (km s$^{-1}$) & \\
\hline
G075.29+01.32& 20144+3726 &      &      & 10.75$\pm$0.50 & 0.22$\pm$0.01 &12.0$\pm$ 6.0&   0.4$\pm$9.4& -17.9$\pm$ 6.0&1 \\
G097.53+03.18& 21306+5540 &91    &S128N & 11.90$\pm$1.52 & 0.42$\pm$0.05 &10.8$\pm$3.6&   1.1$\pm$14.4&   8.9$\pm$ 4.1&3 \\
G135.27+02.79& 02395+6244 &437   &      & 13.24$\pm$0.87 & 0.29$\pm$0.02 &14.8$\pm$5.3&   2.5$\pm$ 8.4&   0.9$\pm$9.9&2 \\
G168.06+00.82& 05137+3919 &621   &      & 13.21$\pm$1.58 & 0.07$\pm$0.01 &9.8$\pm$ 5.3& -11.3$\pm$ 7.9&   4.8$\pm$ 7.5&3\\
G182.67--03.26&05363+2454 &      &      & 14.66$\pm$3.92 &-0.36$\pm$0.10 &13.1$\pm$10.0& --5.7$\pm$12.3&  11.1$\pm$11.0&3 \\
\hline
\end{tabular} \\ 
\end{center}
{\footnotesize 
References:(1) Sanna et al. 2012, (2) Hachisuka et al. 2009 and (3) this paper.
$(U_s,V_s,W_s)$ are the peculiar motion components toward the Galactic center, in the
direction of Galactic rotation,  and toward the north Galactic pole. 
}
\end{table*}

\subsection{Structure of the Outer arm}

\begin{figure}[ht]
\resizebox{\hsize}{!}{\includegraphics[angle=0]{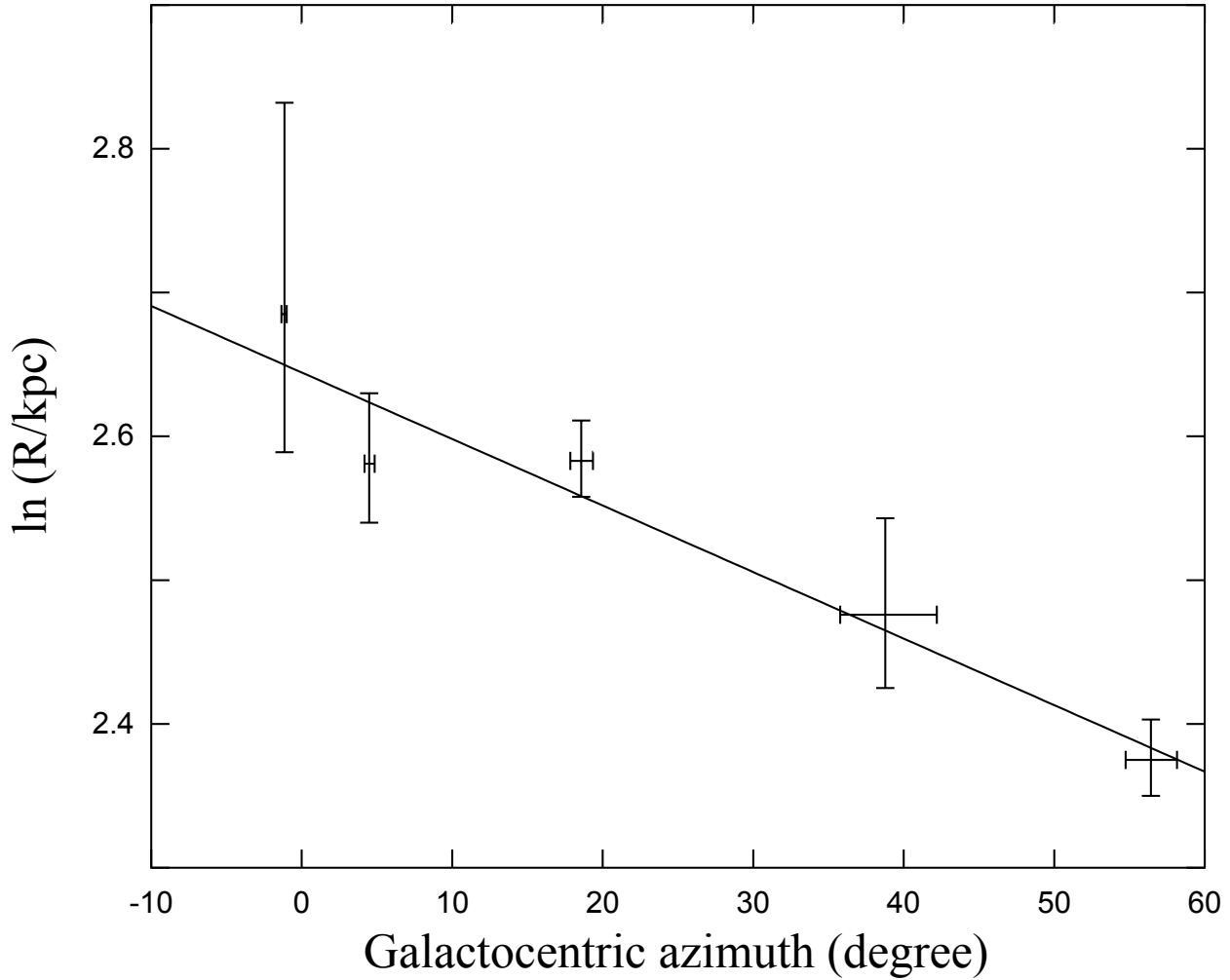}}
\caption{
{\footnotesize
Galactocentric distance vs. azimuth. The line is the 
result of log-periodic spiral fitting. 
}
}
\label{pitch}
\end{figure}

The Outer Arm as it passes through the second and third Galactic quadrants should
be close to the outer ``edge'' of the spiral structure as traced by stars,
and its location may mark the outer range of active star formation in the Milky Way. 
With six sources in the Outer Arm, spanning a large range
of Galactocentric azimuth, $\beta$ (the angle between the Sun and a source as viewed
from the Galactic center and increasing with Galactic longitude), 
we estimate the pitch angle by fitting a section of a log-periodic spiral pattern 
to $\ln(R/{\rm kpc})$ versus $\beta$, as described in Reid et al (2014).   
The data are plotted in Figure~\ref{pitch} and the pitch angle is estimated to be 
$14\d9 \pm 2\d7$. 
This estimate is consistent within 1$\sigma$ with the value reported in  Reid et al (2014), 
which included the contribution of the star forming region G196.45-01.67 as well. 
Also, our estimate agrees within 2$\sigma$-3$\sigma$ with the range of pitch angles reported by 
Hou \& Han (2014), who made use of a compilation of different star formation tracers

The Galactocentric distance of Outer Arm at $\beta=0$ (toward the
Galactic anticenter) is $14.1\pm0.6$ kpc.  This is 
consistent with the size of stellar disk estimated from  
red-clump giant stars of $13.9\pm0.5$ kpc by Minniti et al. (2011),
provided the Outer Arm dies off before reaching much greater distance.  

\begin{figure}[ht]
\resizebox{\hsize}{!}{\includegraphics[angle=0]{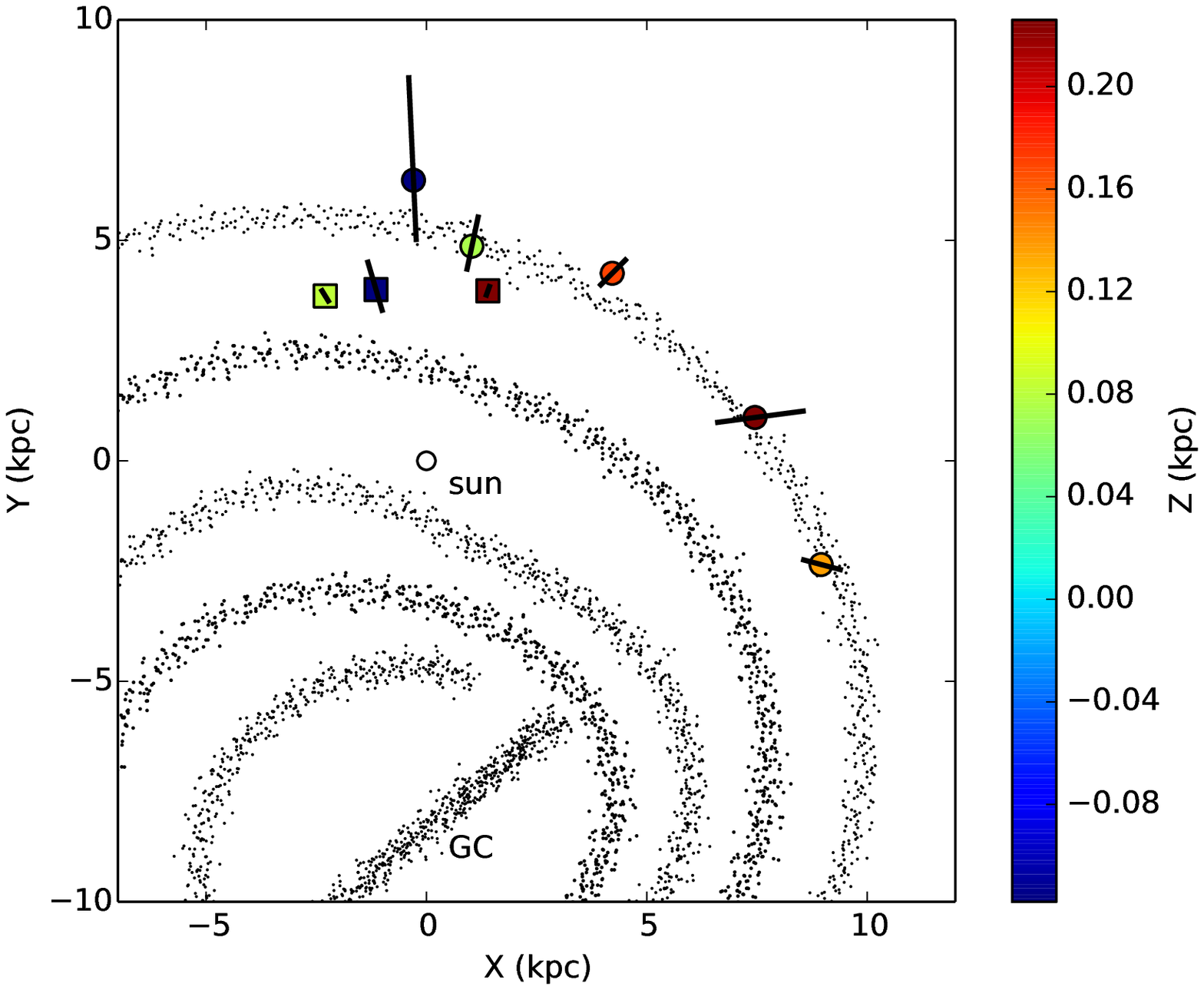}}
\caption{
Location of 22 GHz water maser sources in the Outer Arm (filled circle) with error bar. 
Other 22 GHz water maser sources in the Outer Arm, G075.29+01.32 (Sanna et al. 2012) 
and G135.27+02.79 (Hachisuka et al. 2009),  are also plotted (filled circle). 
Three 22 GHz water maser sources in the interarm (filled square), G160.14+03.15,  
G211.59+01.05 (Reid et al. 2014), and G196.45-01.67 (Asaki et al. 2014) are plotted.
The color index shows the distance from the Galactic plane. The origin of 
figure is the location of Sun. The distance between Sun and Galactic center is 8.3 kpc. 
}
\label{z}
\end{figure}

Because the Outer Arm from the second through the fourth Galactic quadrants is at
a great distance from the Galactic center, the arm is subject to possible 
warping through interactions with other galaxies in the Local Group (e.g., Purcell et al. 2011). 
Warping is clearly seen for HI gas beyond the stellar disk (e.g., Levine et al. 2006). 
The offsets of the star-forming regions with parallax distances (spanning
$75^\circ < l < 183^\circ $) perpendicular to the 
Galactic plane show clear signs of warping (see Table~\ref{table:4-1} and  Figure~\ref{z}).  
Urquhart et al (2014) suggest that the mean amplitude of warping for red MSX
sources at a Galactocentric distance of 11.5 kpc is about 160 pc from the Galactic plane ($b=0^{\circ}$), and
our data show a warping reaching 400 pc at 13 kpc. 
This tendency is consistent with the warping of the HI disk of the Milky Way beyond
13 kpc noted by Levine et al. (2006), Kalberla et al. (2007), and Kalberla \& Kerp (2009).

The annual parallax of 0.189$\pm$0.008 mas for water maser source G$196.45-01.67$ 
(IRAS 06117+1350 or S269) was measured by VLBI Exploration of Radio Astrometry (VERA) 
using one maser spot; this large distance suggested that the maser source was 
a part of the Outer Arm (Honma et al. 2007).  Recently, Asaki et al (2014) re-analyzed
the VERA data for this source using a more compact maser spot, and obtained a parallax  
of 0.247$\pm$0.034 mas.  The revised VERA parallax distances suggests that G$196.45-01.67$ 
is closer to the Sun than we find by spiral arm fitting of five Outer Arm sources.
Therefore, G$196.45-01.67$ might be an interam source between Perseus and Outer Arms.  
Indeed, two other maser sources toward the general anticenter direction found by the BeSSeL 
Survey (Reid et al. 2014) also seem to lie between the Perseus and Outer Arms (Figure~\ref{z}).
Possibly, the Outer Arm splits into two branches in this region;  
more astrometric results are needed to clarify this issue.

\subsection{Velocity Field in the Outer Arm}

The Galactic rotation speed in the outer Galaxy is an important parameter
for estimating the distribution of mass among the Galactic disk, bulge, and halo.
Assuming the Galactic and Solar motion parameters given above,
we find small peculiar motions in the direction of Galactic rotation 
($V_s$ in Table~\ref{table:4-1}) for these sources.
The weighted average for $\Vsbar$ values is $-3.3\pm4.4$ \kms. 
This suggests that the Galactic rotation speed remains reasonably constant to 
Galactocentric distances of at least 15 kpc, supporting the existence of a 
dark matter halo. Although the component of peculiar motions toward the north 
Galactic pole (\Ws\ in Table~\ref{table:4-1}) display some scatter,  
the weighted average ($\Wsbar$) of these values is $1.8\pm2.9$ \kms,  
which suggests that the maser sources on average are moving almost entirely 
in the Galactic plane. 

It is interesting to note that all five maser sources clearly identified with
the Outer Arm have peculiar motion components toward the Galactic center 
(i.e., positive \Us\ values in Table~\ref{table:4-1}),
suggesting that the sources in the Outer Arm are not in simple circular Galactic 
orbits.  The weighted average ($\Usbar$) is $11.7\pm2.3$ \kms.

\subsection{Other Maser Sources in the Outer Srm}

In order to better understand the structure and dynamics of outer Galaxy will require
more astrometric data.  Surveys for 6.7 GHz methanol maser sources beyond the Perseus Arm have been performed, but only a few sources have been found (Pestalozzi et al. 2005, Xu et al. 2008). Of these sources, two (G$097.53+03.18$ and G$168.06+00.82$) belong 
to star forming regions
with 22 GHz water masers reported in this paper, and a third (G$196.45-01.67$) 
displays both methanol and water masers (Menten 1991; Rygl et al. 2010) and 
an astrometric result has been reported (Honma et al. 2007; Asaki et al 2014).   

Although extensive surveys for 22 GHz water masers have been performed, 
only a few tens of sources have been found with Galactocentric distances that might be beyond 
the Perseus arm ($\gax 10$ kpc; Woulterloot et al. 1993, 1988).
Unfortunately, most outer Galaxy water masers are weak and upcoming parallax observations
for sources in the Outer Arm will require greater sensitivity.  This can be
provided by the VLBA with its recent bandwidth upgrade, which allows the use of
weaker background continuum sources as phase calibrators.

\section{Summary}

We have performed VLBA parallax and proper motion observations of three water 
maser sources associated with regions of high-mass star formation in the Outer Arm
of the Milky Way.  These observations are part of the BeSSeL 
Survey, an NRAO Key Science Project.  Combining our results with published results, 
we find that the Galactocentric distance of the Outer Arm is 
$14.1 \pm 0.6$ kpc in the direction of the anticenter, which is consistent with estimates of the 
outer edge of the stellar disk.   The pitch angle of the Outer Arm in the second
and third Galactic quadrants is $14\d9 \pm 2\d7$.
Our findings, though based on a limited number of sources, 
indicate a tendency for peculiar motions toward the Galactic center,
as might be expected for gas entering a trailing spiral arm and being shock and then
forming stars.  In order to better determine the structure
of the Outer Arm, further astrometric observation of weak water maser sources 
are needed.

\acknowledgments
This work is partly supported by China Ministry of Science and Technology 
under State Key Development Program for Basic Research (2012CB821800), 
the National Natural Science Foundation of China (grants 10625314, 11121062).
The authors acknowledge the financial support by the European Research Council for the ERC Advanced Grant GLOSTAR under contract no. 247078.

{\it Facilities:} \facility{VLBA}

\section{Online Material}

\begin{table*}
\scriptsize
\begin{center}
\caption{Results of Parallax and proper motion measurements.\label{table:g097a}}
\begin{tabular}{cccccc}
\hline \hline 
Maser        & Background & $v_{\rm LSR}$ & Parallax          & $\mu_{\alpha}$  & $\mu_{\delta}$   \\
             & source     & (km s$^{-1}$) & (mas)             & (mas yr$^{-1}$) & (mas yr$^{-1}$)  \\
\hline
G097.53+3.18 & J2127+5528   & --69.42     & 0.126 $\pm$ 0.027 & --2.69 $\pm$ 0.13 & --2.88 $\pm$ 0.07 \\
             & J2139+5540   & --69.42     & 0.171 $\pm$ 0.023 & --2.71 $\pm$ 0.13 & --2.80 $\pm$ 0.06 \\
             & J2123+5500   & --69.42     & 0.141 $\pm$ 0.018 & --2.66 $\pm$ 0.05 & --3.00 $\pm$ 0.13 \\
             & J2117+5431   & --69.42     & 0.118 $\pm$ 0.016 & --2.62 $\pm$ 0.04 & --3.03 $\pm$ 0.12 \\
             & Combined fit & --69.42     & 0.120 $\pm$ 0.014 & --2.67 $\pm$ 0.10 & --2.93 $\pm$ 0.10 \\

G097.53+3.18 & J2127+5528   & --69.84     & 0.066 $\pm$ 0.028 & --3.00 $\pm$ 0.09 & --2.74 $\pm$ 0.08 \\
             & J2139+5540   & --69.84     & 0.117 $\pm$ 0.030 & --3.06 $\pm$ 0.11 & --2.79 $\pm$ 0.07 \\
             & J2123+5500   & --69.84     & 0.074 $\pm$ 0.029 & --2.93 $\pm$ 0.08 & --2.80 $\pm$ 0.09 \\
             & J2117+5431   & --69.84     & 0.065 $\pm$ 0.029 & --2.87 $\pm$ 0.08 & --2.78 $\pm$ 0.14 \\
             & Combined fit & --69.84     & 0.070 $\pm$ 0.014 & --2.96 $\pm$ 0.09 & --2.78 $\pm$ 0.10 \\

G097.53+3.18 & J2127+5528   & --71.52     & 0.213 $\pm$ 0.015 & --3.60 $\pm$ 0.05 & --2.34 $\pm$ 0.22 \\
             & J2139+5540   & --71.52     & 0.201 $\pm$ 0.029 & --3.57 $\pm$ 0.10 & --2.35 $\pm$ 0.11 \\
             & J2123+5500   & --71.52     & 0.153 $\pm$ 0.023 & --3.19 $\pm$ 0.07 & --2.32 $\pm$ 0.26 \\
             & J2117+5431   & --71.52     & 0.122 $\pm$ 0.038 & --3.09 $\pm$ 0.12 & --2.13 $\pm$ 0.29 \\
             & Combined fit & --71.52     & 0.169 $\pm$ 0.017 & --3.36 $\pm$ 0.09 & --2.29 $\pm$ 0.23 \\

G097.53+3.18 & J2127+5528   & --71.94     & 0.221 $\pm$ 0.014 & --3.63 $\pm$ 0.04 & --2.38 $\pm$ 0.22 \\
             & J2139+5540   & --71.94     & 0.214 $\pm$ 0.030 & --3.61 $\pm$ 0.09 & --2.40 $\pm$ 0.13 \\
             & J2123+5500   & --71.94     & 0.162 $\pm$ 0.018 & --3.23 $\pm$ 0.06 & --2.36 $\pm$ 0.26 \\
             & J2117+5431   & --71.94     & 0.132 $\pm$ 0.034 & --3.13 $\pm$ 0.10 & --2.18 $\pm$ 0.30 \\
             & Combined fit & --71.94     & 0.177 $\pm$ 0.016 & --3.40 $\pm$ 0.08 & --2.33 $\pm$ 0.24 \\

G097.53+3.18 & J2127+5528   & --72.37     & 0.211 $\pm$ 0.010 & --3.63 $\pm$ 0.03 & --2.38 $\pm$ 0.22 \\
             & J2139+5540   & --72.37     & 0.206 $\pm$ 0.025 & --3.61 $\pm$ 0.08 & --2.41 $\pm$ 0.13 \\
             & J2123+5500   & --72.37     & 0.150 $\pm$ 0.021 & --3.21 $\pm$ 0.06 & --2.36 $\pm$ 0.26 \\
             & J2117+5431   & --72.37     & 0.116 $\pm$ 0.040 & --3.11 $\pm$ 0.12 & --2.17 $\pm$ 0.30 \\
             & Combined fit & --72.37     & 0.166 $\pm$ 0.016 & --3.39 $\pm$ 0.08 & --2.33 $\pm$ 0.24 \\

G097.53+3.18 & J2127+5528   & --77.00     & 0.117 $\pm$ 0.018 & --3.02 $\pm$ 0.05 & --2.34 $\pm$ 0.06 \\
             & J2139+5540   & --77.00     & 0.145 $\pm$ 0.016 & --3.09 $\pm$ 0.05 & --2.37 $\pm$ 0.05 \\
             & J2123+5500   & --77.00     & 0.164 $\pm$ 0.013 & --2.95 $\pm$ 0.03 & --2.42 $\pm$ 0.14 \\
             & J2117+5431   & --77.00     & 0.147 $\pm$ 0.018 & --2.89 $\pm$ 0.04 & --2.40 $\pm$ 0.16 \\
             & Combined fit & --77.00     & 0.140 $\pm$ 0.009 & --2.99 $\pm$ 0.04 & --2.42 $\pm$ 0.11 \\

G097.53+3.18 & J2127+5528   & --77.42     & 0.120 $\pm$ 0.019 & --3.03 $\pm$ 0.05 & --2.38 $\pm$ 0.07 \\
             & J2139+5540   & --77.42     & 0.147 $\pm$ 0.016 & --3.09 $\pm$ 0.05 & --2.41 $\pm$ 0.05 \\
             & J2123+5500   & --77.42     & 0.167 $\pm$ 0.013 & --2.95 $\pm$ 0.03 & --2.46 $\pm$ 0.14 \\
             & J2117+5431   & --77.42     & 0.150 $\pm$ 0.018 & --2.89 $\pm$ 0.04 & --2.44 $\pm$ 0.16 \\
             & Combined fit & --77.42     & 0.143 $\pm$ 0.009 & --2.99 $\pm$ 0.04 & --2.42 $\pm$ 0.11 \\
G097.53+3.18 & J2127+5528   & --77.84     & 0.209 $\pm$ 0.017 & --2.97 $\pm$ 0.04 & --2.32 $\pm$ 0.08 \\
             & J2139+5540   & --77.84     & 0.237 $\pm$ 0.016 & --3.03 $\pm$ 0.05 & --2.36 $\pm$ 0.05 \\
             & J2123+5500   & --77.84     & 0.242 $\pm$ 0.029 & --2.89 $\pm$ 0.07 & --2.40 $\pm$ 0.15 \\
             & J2117+5431   & --77.84     & 0.230 $\pm$ 0.030 & --2.83 $\pm$ 0.08 & --2.38 $\pm$ 0.17 \\
             & Combined fit & --77.84     & 0.228 $\pm$ 0.012 & --2.93 $\pm$ 0.06 & --2.37 $\pm$ 0.12 \\
             
G097.53+3.18 & J2127+5528   & --78.26     & 0.122 $\pm$ 0.018 & --2.37 $\pm$ 0.05 & --2.41 $\pm$ 0.10 \\
             & J2139+5540   & --78.26     & 0.149 $\pm$ 0.016 & --2.43 $\pm$ 0.04 & --2.44 $\pm$ 0.06 \\
             & J2123+5500   & --78.26     & 0.165 $\pm$ 0.016 & --2.29 $\pm$ 0.04 & --2.49 $\pm$ 0.16 \\
             & J2117+5431   & --78.26     & 0.148 $\pm$ 0.022 & --2.23 $\pm$ 0.06 & --2.47 $\pm$ 0.19 \\
             & Combined fit & --78.26     & 0.144 $\pm$ 0.010 & --2.33 $\pm$ 0.05 & --2.45 $\pm$ 0.14 \\
G097.53+3.18 & J2127+5528   & --78.69     & 0.090 $\pm$ 0.020 & --2.96 $\pm$ 0.05 & --2.23 $\pm$ 0.10 \\
             & J2139+5540   & --78.69     & 0.124 $\pm$ 0.019 & --3.03 $\pm$ 0.05 & --2.26 $\pm$ 0.06 \\
             & J2123+5500   & --78.69     & 0.129 $\pm$ 0.011 & --2.89 $\pm$ 0.03 & --2.31 $\pm$ 0.13 \\
             & J2117+5431   & --78.69     & 0.112 $\pm$ 0.016 & --2.83 $\pm$ 0.04 & --2.28 $\pm$ 0.17 \\
             & Combined fit & --78.69     & 0.109 $\pm$ 0.009 & --2.93 $\pm$ 0.04 & --2.27 $\pm$ 0.12 \\
G097.53+3.18 & J2127+5528   & --79.53     & 0.106 $\pm$ 0.017 & --2.46 $\pm$ 0.04 & --2.66 $\pm$ 0.07 \\
             & J2139+5540   & --79.53     & 0.133 $\pm$ 0.016 & --2.53 $\pm$ 0.04 & --2.69 $\pm$ 0.06 \\
             & J2123+5500   & --79.53     & 0.146 $\pm$ 0.015 & --2.39 $\pm$ 0.04 & --2.74 $\pm$ 0.13 \\
             & J2117+5431   & --79.53     & 0.130 $\pm$ 0.020 & --2.33 $\pm$ 0.05 & --2.72 $\pm$ 0.15 \\
             & Combined fit & --79.53     & 0.126 $\pm$ 0.009 & --2.43 $\pm$ 0.04 & --2.70 $\pm$ 0.11 \\
\hline
             & Combined fit &             & 0.133 $\pm$ 0.017 &                   &                   \\
             & $<\mu>$      & --74.89     &                   & --2.94 $\pm$ 0.07 & --2.48 $\pm$ 0.16 \\
\hline
\end{tabular}
\end{center}
\end{table*}

\begin{table*}
\begin{center}
\caption{Results of Parallax and proper motion measurements. \label{table:g168}}
\begin{tabular}{cccccc}
\hline \hline 
Maser        & Background & $v_{\rm LSR}$ & Parallax          & $\mu_{\alpha}$  & $\mu_{\delta}$   \\
             & source     & (km s$^{-1}$) & (mas)             & (mas yr$^{-1}$) & (mas yr$^{-1}$)  \\
\hline
G168.06+00.82& J0512+4041   & --26.21    & 0.210 $\pm$ 0.019 & 0.77 $\pm$ 0.05 & --0.07 $\pm$ 0.04 \\
                         & J0523+3926   & --26.21     & 0.249 $\pm$ 0.028 & 0.74 $\pm$ 0.08 & 0.04 $\pm$ 0.06 \\
                         & J0509+3951   & --26.21    & 0.239 $\pm$ 0.028 & 0.85 $\pm$ 0.08 & --0.05 $\pm$ 0.11 \\
                         & Combined fit  & --26.21     & 0.230 $\pm$ 0.024 & 0.79 $\pm$ 0.07 & --0.03 $\pm$ 0.08 \\
G168.06+00.82& J0512+4041   & --28.74     & 0.177 $\pm$ 0.015 & 0.47 $\pm$ 0.04 & --1.10 $\pm$ 0.05 \\
                         & J0523+3926   & --28.74     & 0.210 $\pm$ 0.016 & 0.43 $\pm$ 0.05 & --0.99 $\pm$ 0.03 \\
                         & J0509+3951   & --28.74    & 0.204 $\pm$ 0.017 & 0.55 $\pm$ 0.05 & --1.08 $\pm$ 0.08 \\
                         & Combined fit & --28.74    & 0.194 $\pm$ 0.016 & 0. 48$\pm$ 0.05 & --1.06 $\pm$ 0.06 \\
G168.06+00.82& J0512+4041   & --29.16     & 0.193 $\pm$ 0.026 & 0.58 $\pm$ 0.07 & --1.00 $\pm$ 0.08 \\
                         & J0523+3926   & --29.16     & 0.222 $\pm$ 0.032 & 0.54 $\pm$ 0.09 & --0.89 $\pm$ 0.09 \\
                         & J0509+3951   & --29.16     & 0.219 $\pm$ 0.032 & 0.66 $\pm$ 0.09 & --0.98 $\pm$ 0.16 \\
                         & Combined fit & --29.16     & 0.211 $\pm$ 0.029 & 0. 59$\pm$ 0.08 & --0.96$\pm$ 0.12 \\
G168.06+00.82& J0512+4041   & --29.58     & 0.156 $\pm$ 0.013 & 0.46 $\pm$ 0.04 & --1.14 $\pm$ 0.07 \\
                         & J0523+3926   & --29.58     & 0.182 $\pm$ 0.019 & 0.42 $\pm$ 0.05 & -1.02 $\pm$ 0.03 \\
                         & J0509+3951   & --29.58     & 0.179 $\pm$ 0.022 & 0.53 $\pm$ 0.06 & --1.12 $\pm$ 0.07 \\
                         & Combined fit & -- 29.58    & 0.170 $\pm$ 0.019 & 0.47 $\pm$ 0.05 & --1.09 $\pm$ 0.06 \\
\hline 
             & Combined fit &             & 0.201 $\pm$ 0.024 &                 &                   \\
             & $<\mu>$      & --28.42     &                   & 0.58 $\pm$ 0.07 & --0.78 $\pm$ 0.08 \\ 
\hline 
\end{tabular}
\end{center}
\end{table*}

\begin{table*}
\begin{center}
\caption{Results of Parallax and proper motion measurements.\label{table:g182}}
\begin{tabular}{cccccc}
\hline \hline 
Maser        & Background & $v_{\rm LSR}$ & Parallax          & $\mu_{\alpha}$  & $\mu_{\delta}$   \\
             & source     & (km s$^{-1}$) & (mas)             & (mas yr$^{-1}$) & (mas yr$^{-1}$)  \\
\hline
G182.68--03.26 & J0540+2507   & --5.37     & 0.158 $\pm$ 0.045 & 0.74 $\pm$ 0.12 & --0.44 $\pm$ 0.14 \\
& J0540+2507   & --5.79     & 0.149 $\pm$ 0.042 & 0.74 $\pm$ 0.11 & --0.45 $\pm$ 0.13 \\
& J0540+2507   & --7.05     & 0.191 $\pm$ 0.042 & 0.01 $\pm$ 0.11 & --0.59 $\pm$ 0.14 \\
& J0540+2507   & --7.89     & 0.131 $\pm$ 0.044 & 0.26 $\pm$ 0.12 & --0.30 $\pm$ 0.12 \\

\hline 
             & Combined fit &             & 0.157 $\pm$ 0.042 &                 &                   \\
             & $<\mu>$      & --6.53     &                   & 0.44 $\pm$ 0.12 & --0.45 $\pm$ 0.13 \\ 
\hline 

\hline
\end{tabular}
\end{center}
\end{table*}

\end{document}